\documentclass[a4paper,10pt]{article}
\usepackage{amssymb} 
\usepackage{amsmath} 
\usepackage{amsthm}
\font\frak=eufm10 scaled\magstep1
\def\goth #1{\hbox{{\frak #1}}}
\def\pd#1#2{\frac{\partial#1}{\partial#2}}
\def\matriz#1#2{\left( \begin{array}{#1} #2 \end{array}\right) }
\def\GR{{\mathcal{G}}}

\title{Lie systems and integrability conditions of differential equations\\
and some of its applications}
\author{J. F. CARI\~NENA$^*$ and J. de LUCAS$^{**}$}

\begin{document}
\maketitle

\centerline{Departamento de F\'{\i}sica Te\'orica, Facultad de Ciencias,
   Universidad de Zaragoza,}
\centerline{50009 Zaragoza, Spain}
\centerline{$^*$E-mail: jfc@unizar.es,\quad $^{**}$E-mail: dlucas@unizar.es}

\begin{abstract}
The geometric theory of Lie systems is used to establish
 integrability conditions for several systems of differential equations, in
 particular some Riccati
equations and Ermakov systems. Many different integrability criteria in the
literature will be analysed from this new perspective, and some applications in physics
will be given.
\end{abstract}

\section{Introduction}

Non-autonomous systems of first-order and second-order differential equations appear in many
places in physics. For instance, 
Hamilton equations are  systems of first-order differential  equations, while
Euler--Lagrange 
equations for regular Lagrangians are systems of second-order
differential  equations.
A system of second-order differential equations in $n$ variables of the form $\ddot x^i=F^i(x,\dot x,t)$, with $i=1,\ldots,n$, is related with a system of
first-order equations in $2n$ variables:
\begin{equation}
\left\{\begin{array}{rcl} \dot
    x^i&=&v^i\\ \dot v^i&=&F^i(x,v,t)\end{array}\right. \,,\qquad i=1,\ldots n\,.\label{nonasodesys}
\end{equation}
Therefore, it is enough to restrict ourselves to study systems of  first-order 
differential equations.
 From the geometric viewpoint a system
\begin{equation}
\dot x^i=X^i(x,t)\,,\qquad i=1,\ldots,n\,,\label{nonasys}
\end{equation}
is associated with the $t$-dependent vector field 
$X= X^i(x,t)\,\partial/\partial {x^i}$
whose integral curves are determined by the solutions of the system.

Unfortunately, there is no general method for solving such equations.
Relevant questions about integrability are 
 how to find a particular solution (determined by $x(0)=x_0)$, or
 a $r$-parameter family of solutions, or even 
 the general solution (a $n$-parameter family of solutions). Finally, when
is it possible to find  and how to determine  a superposition rule for  solutions?

We shall understand that to find a solution means to reduce the problem  to
carry out some quadratures. For instance,
the general solution of the inhomogeneous linear differential equation
${dx}/{dt}=b_0(t)+b_1(t)x$
can be found with  two quadratures and it is 
given by 
$$
x(t)=\exp\left(\int_0^tb_1(s)\,ds\,\right) \times \left(x_0+\int_0^t b_0(t^{\prime})
        \exp\left(- \int_0^{t^{\prime}} b_1(s) \,ds \right) dt^{\prime}
\right)\,.
$$

Actually, when the systems we are dealing with are linear, there is a {\sl
  linear
superposition principle}
allowing us to find the general solution as a linear combination of $n$
particular solutions.
For instance,  for the harmonic oscillator with a $t$-dependent angular frequency
$\omega(t)$:
$$ 
\ddot x= -\omega^2(t)\, x
\Longleftrightarrow \left\{\begin{array}{rcl}\dot x&=&v\cr \dot
    v&=&-\omega^2(t)\, x\end{array}\right.,
$$
whose solutions are the integral curves of the 
$t$-dependent vector field 
$X=v\partial/\partial x -\omega^2(t) x\,\partial/\partial v$,
if we know a particular solution, the general solution can be found by
means of one quadrature and  if we know two particular solutions, $x_1$ and $x_2$, 
the general solution is a linear
combination (no quadrature is needed) 
$x(t)=k_1\,x_1(t)+k_2\, x_2(t)$.

There are systems whose general solution can be written as a
nonlinear function of some particular solutions. For instance, for  Riccati equation: 
if a particular solution is known, the general solution is obtained by two
quadratures, if two particular solutions are known the problem reduces to 
one quadrature and, finally, when three particular solutions 
are known,  $x_1,x_2$ and  $x_3$, the general solution can be found from 
the cross-ratio relation
$$
\frac{x-x_1}{x-x_2}:\frac{x_3-x_1}{x_3-x_2}=k\ ,
$$
which provides us a nonlinear superposition rule\cite{CarRam}.

There also exist cases in which we can superpose solutions of one system for
finding solutions of another one. We have seen one example:
the general solution of the inhomogeneous linear equation can be written as 
$x(t)=x_1(t)+C\, x_0(t)$,
where
$x_0(t)$ is a solution of the associated homogeneous equation
and $x_1(t)$ is a particular solution of the inhomogeneous linear one.

Milne-Pinney equation\cite{Pi50} $\ddot x=-\omega^2(t)x+ k/{x^3}$ is usually studied
together with the time-dependent harmonic oscillator 
 $\ddot y+\omega^2(t) y=0$ and the system is called Ermakov system.
 Pinney showed  in a short
 paper \cite{Pi50}  that the general solution of the first equation 
can be written as a nonlinear superposition of  two solutions of the associated
harmonic oscillator. 
All these properties can be better understood in the framework of Lie systems,
conveniently extended in some cases to include systems of second-order differential
equations.
These systems have a lot of applications not only in mathematics but also in
many different branches of classical and quantum physics. 

 Let us look for systems admitting a (maybe nonlinear) superposition rule.
The main result was given by Lie \cite{LS,{CGM00}}:

{\bf Theorem:} \ 
{\it Given (\ref{nonasys}) 
a necessary and sufficient condition for the existence of a function 
$\Phi:{\mathbb{R}}^{n(m+1)}\to {\mathbb{R}}^n$ such that the general solution is
$x=\Phi(x_{(1)}, \ldots,x_{(m)};k_1,\ldots,k_n)$, 
with $\{x_{(a)}\mid a=1,\ldots,m\}$
being a set of particular solutions of the system
and  $k_1,\ldots,k_n,$ 
are $n$  arbitrary constants,
is that the system can be written as  
$$\frac {dx^i}{dt}=Z^1(t)\xi_1^{i}(x)+\cdots+Z^r(t)\xi_r^{i}(x), $$
where  $Z^1,\ldots,Z^r,$ are  $r$ functions  depending only on  $t$ and
  $\xi_{\alpha}^i$, $\alpha=1,\ldots, r$,  are functions of 
 $x=(x^1,\ldots,x^n)$,
such that the  $r$ vector fields in   ${\mathbb{R}}^n$ given by
$X_{\alpha}\equiv   {\displaystyle\sum_{i=1}^n}\xi_\alpha^{i}(x^1,\ldots,x^n)\partial/\partial {x^i}$,
$\alpha=1,\ldots,r$,
close on a real  finite-dimensional Lie algebra, i.e. the $X_\alpha$ are l.i. 
and  there are $r^3$
real numbers, $c_{\alpha\beta}\,^\gamma$, such that
 $[X_\alpha,X_\beta]={\displaystyle\sum_{\gamma=1}^r}c_{\alpha\beta}\,^\gamma X_\gamma$. 
The number  $r$ satisfies $r\leq mn$.}

The condition in the Theorem is that $X(x,t)$  can be written as 
$X(x,t)={\displaystyle\sum_{\alpha=1}^r}Z^\alpha(t)X_\alpha(x)$,
with  $X_\alpha$ as mentioned above.

Non-autonomous systems corresponding to such $t$-dependent vector fields
 will be
called Lie systems.
One instance of Lie system is
the Riccati equation\cite{CarRam}
\begin{equation}
\frac{dx(t)}{dt}=b_2(t)\,x^2(t)+b_1(t)\,x(t)+ b_0(t)\ ,\label{Ricceq}
\end{equation}
for which $m=3$ and the superposition principle comes from the relation 
$$
\frac{x-x_1}{x-x_2}:\frac{x_3-x_1}{x_3-x_2}=k\ \Longrightarrow 
x=\frac {k\, x_1(x_3-x_2)+x_2(x_1-x_3)}{k\,(x_3-x_2)+(x_1-x_3)}\ .
$$
The associated Lie algebra is generated by  $X_{0}$, $X_{1}$
 and $X_{2}$ given by 
$$
X_{0} =\pd{}{x}\,,        \quad
X_{1} =x\,\pd{}{x}\, ,    \quad
X_{2} = x^2\,\pd{}{x}\,,  
$$
which  close on a ${\goth{sl}}(2,{\mathbb{R}})$ 3-dimensional real Lie  algebra,
because
\begin{equation}
[X_{0},X_{1}] = X_{0} \,,      \quad 
[X_{0},X_{2}] = 2X_{1}\,,     \quad
[X_{1},X_{2}] = X_{2} \,.\label{comsl2}
\end{equation}

The time-dependent harmonic oscillator is also an example of physical relevance.
It is described by a
Hamiltonian 
$$
H=\frac 12\,\frac {p^2}{m(t)}+\frac 12\, m(t)\omega^2(t)x^2\,,
$$
which gives rise to the dynamics defined by 
the $t$-dependent vector field 
$$X(x,p,t)=\frac 1{ m(t)}\,p\,\pd{}x-m(t)\omega^2(t)\,x\pd{}p\,.
$$
If we consider the set of vector fields
$$
X_0=p\pd{}{x},\quad X_1=\frac 12\left(x\pd{}{x}-p\pd{}{p}\right),\quad X_2=-x\pd{}{p}\,,
$$
which close on a $\goth{sl}(2,\mathbb{R})$ Lie algebra with the same 
 commutation relations as (\ref{comsl2}),
 the corresponding $t$-dependent vector field $X$ 
can be written as a linear combination
$ X(\cdot,t)=m(t)\omega^2(t)\,X_2(\cdot)+(1/m(t))\,X_0(\cdot)$,
i.e. it is a linear combination with $t$-dependent coefficients 
$
X(\cdot,t)={\displaystyle\sum_{\alpha=0}^2}b_\alpha(t)X_\alpha(\cdot)$
 with 
 $b_0(t)=1/m(t)$, $b_1(t)=0$ and $ b_2(t)=m(t)\omega^2(t)$.

The prototype of Lie system is the time-dependent right-invariant vector fields in a Lie group $G$.
Let   $\{a_1,\ldots,a_r\}$ denote 
a basis of $T_eG$. A  right-invariant vector field $X^R$ is one such that 
$X^R(g)=R_{g*e}X^R(e)$. Define $X^R_\alpha$ by $X^R_\alpha(g)=R_{g*e}a_\alpha$.
The $t$--dependent right-invariant vector field 
$$\bar X(g,t)=-\sum_{\alpha=1}^r b_\alpha(t)X^R_\alpha(g)\ ,
$$
defines a Lie system in $G$ whose integral curves 
are  solutions of the system
$\dot g=-\sum_{\alpha=1}^r b_\alpha(t)\,X^R_\alpha(g)$,
and when applying $R_{g^{-1}}$ to both sides we see that $g(t)$ satisfies
\begin{equation}
R_{g^{-1}(t)*g(t)}\dot g(t)\,=-\sum_{\alpha=1}^r b_\alpha(t)a_\alpha\in T_eG
\ .\label{lieingr}
\end{equation}
Let $H$ be a closed subgroup of $G$ and consider the homogeneous space $M=G/H$.
Then, $G$ can be seen as a principal bundle over $G/H$: $(G,\tau,G/H)$.
The 
$X^R_\alpha$
are $\tau$-projectable on  the corresponding fundamental vector fields of the left-action
$\lambda:(g,g'H)\in G\times M \rightarrow (gg'H)\in M$ given by  
$-X_\alpha=-
X_{a_\alpha}$ with 
$\tau_{*g}X_\alpha^R(g)=-X_\alpha(gH)$,
 the projected  vector field in $M$ will be 
$X(x,t)=\sum_{\alpha=1}^rb_\alpha(t)X_\alpha(x)$,
and its integral curves are the solutions of the system of differential
equations:
$\dot x=\sum_{\alpha=1}^rb_\alpha(t)X_\alpha(x)$.
The solution of this last system starting from $x_0$ is
$x(t)=\lambda(g(t),x_0)$,
with $g(t)$ being the solution of (\ref{lieingr}) such that $g(0)=e$.
This means that  solving such  Lie system in $G$ we are simultaneously solving 
the corresponding problems in all its homogeneous spaces.

\section{SODE Lie systems}

A system of second order differential equations
 can be studied through the
corresponding system of first-order differential equations as indicated in (\ref{nonasodesys}).
We call SODE Lie systems those for which the associated first-order one 
 is a Lie system, i.e. it can be
written as a linear combination with $t$-dependent coefficients of vector fields
closing on a finite-dimensional real Lie algebra. An example is the 
1-dimensional  harmonic oscillator with time-dependent frequency, but the same
is true for the 2-dimensional isotropic harmonic oscillator with time-dependent 
frequency,
with an associated  vector field 
$$X=v_1\pd{}{x_1} -\omega^2(t) x_1\, \pd{}{v_1}+
v_2\pd{}{x_2} -\omega^2(t) x_2\, \pd{}{v_2}\ ,
$$
which is a linear combination $X=X_2- \omega^2(t)X_1$ with 
$$X_1=x_1 \pd{}{v_1}+x_2 \pd{}{v_2}
\,,\qquad X_2= v_1\pd{}{x_1}+ v_2\pd{}{x_2}\,,
$$
and then they close once again on a Lie algebra
   isomorphic to $\goth{sl}(2,\mathbb{R})$:
\begin{equation} 
[X_1,X_2]=2\, X_3\,, \quad [X_1,X_3]=- X_1 \,,\quad [X_2,X_3]=X_2\,,\label{otrocomsl2}
\end{equation}
with
$$X_3=\frac 12 \left(x_1\pd{}{x_1}-v_1\pd{}{v_1}+x_2\,
\pd{}{x_2}-v_2\pd{}{v_2}
\right) \,.
$$

The search for the superposition rule for a Lie system consists on looking for
enough number of first integrals, independent of the time-dependent coefficients, in an extended space in which we consider
several replicas of the given vector field.

The 2-dimensional case admits an invariant $F$  given by
 the first integral 
$F(x_1,x_2,v_1,v_2)=x_1v_2-x_2v_1$,
which can be seen as a partial superposition rule\cite{CGM07}.  Actually, if $x_1(t)$ is a
solution of the first equation, then we obtain for each real number $k$
 the first-order differential
equation for the variable $x_2$,
$x_1(t)\, {dx_2}/{dt} =k+\dot x_1(t)x_2$,
from where $x_2$ can be found to be given by  
$x_2(t)=k' x_1(t)+k\, x_1(t)\int^t{x_1^{-2}(\zeta)}\,d\zeta$.
With three copies of the same harmonic oscillator, i.e. $X_1$ and
$X_2$ given by 
$$X_1= v_1\pd{}{x_1}+ v_2\pd{}{x_2}+v  \pd{}{x  }\,,\qquad 
 X_2=x_1 \pd{}{v_1}+x_2 \pd{}{v_2}+x\pd {}v\ ,
$$
there exist two independent first integrals
$F_1(x_1,x_2,x,v_1,v_2,v)=xv_1-x_1v$ and
$F_2(x_1,x_2,x,v_1,v_2,v)=xv_2-x_2v$,
from where we obtain the expected superposition rule:
$$x=k_1\, x_1+k_2\, x_2\,,\qquad v=k_1\, v_1+k_2\, v_2\,.
$$

 Another interesting non-linear example is the  Pinney equation, 
the second order non-linear differential  equation:
$$
\ddot x=-\omega^2(t)x+\frac k{x^3}\,,
$$
where $k$ is a constant, with  
associated $t$-dependent vector field
$$X=v\pd{}x+\left(-\omega^2(t)x+\frac k{x^3}\right)\pd{}v\,,
$$
which  is a Lie system because it can be written as 
$X=L_2-\omega^2(t)L_1$,
where
$
L_1:=x\,{\partial}/{\partial v}$ and $L_2= (k /{x^3})\,{\partial}/{\partial v}
+v\,{\partial}/{\partial x}$ 
generate  a three-dimensional real  Lie algebra  isomorphic  to $\goth{sl}(2,\mathbb{R})$
with nonzero defining relations similar to (\ref{otrocomsl2})
with 
$
 L_3=(1/2) \left(x\,{\partial}/{\partial x}-v\,{\partial}/{\partial v}\right)$.

Note that this isotonic oscillator shares with the harmonic one the property of
having a period independent of the energy, i.e. they are isochronous,
 and in the quantum case they have a equispaced spectrum.
The fact that they have the same associated Lie algebra means that they can be
solved simultaneously in the group $SL(2,\mathbb{R})$ by the same  equation 
$$R_{g^{-1}*g}\dot g\, =\omega^2(t)\, a_1-a_2\,, \quad g(0)=e\,.$$

\section{Ermakov systems}
We can consider the 
generalised Ermakov system given by:
$$
\left\{\begin{array}{rcl}
\ddot{x}&=&{\displaystyle\frac{1}{x^3}}f(y/x)-\omega^2(t)x\cr
\ddot{y}&=&{\displaystyle\frac{1}{y^3}}g(y/x)-\omega^2(t)y
\end{array}\right.
$$
which when  $f(u)=k$ and $g(u)=0$ reduces to the 
Ermakov system. 

This system is described by  the $t$-dependent vector field 
$$X=v_x\,\pd{}{x}+v_y\,\pd{}{v_y}+\left(-\omega^2(t)x+\frac 1 {x^3} f( y/
  x)\right)\pd{}{v_x}+\left(-\omega^2(t)y+\frac 1 {y^3} g(y/
    x)\right)\pd{}{v_y}\,,
$$
which can be written as a linear combination 
$X=N_2-\omega^2(t)\, N_1$,
where $N_1$ and $N_2$ are the vector fields 
$$
N_1=x\frac{\partial }{\partial v_x}+y\frac{\partial }{\partial v_y},\quad N_2=v_x\frac{\partial}{\partial x}+
\frac{1}{x^3}f(y/x)\frac{\partial}{\partial v_x}+v_y\frac{\partial}{\partial y}+
\frac{1}{y^3}g( y/x)\frac{\partial}{\partial v_y},
$$
that  generate a three-dimensional real Lie algebra isomorphic to   $\goth{sl}(2,\mathbb{R})$
with a third generator $$N_3=\frac 12\left(x\frac{\partial}{\partial
    x}-v_x\frac{\partial}{\partial v_x}+y\frac{\partial}{\partial
    y}-v_y\frac{\partial}{\partial v_y}\right)\,.$$ 

There exists  a first  integral for the motion, $F:\mathbb{R}^4\rightarrow \mathbb{R}$,
for any $\omega^2(t)$, which  satisfies 
$N_iF=0$ for $i=1,\ldots,3$, but as $[N_1,N_2]=2N_3$ it is enough to impose
$N_1F=N_2F=0$. The condition  $N_1F=0$, 
implies that  there exists a function $\bar F:\mathbb{R}^3\rightarrow \mathbb{R}$ such that
$F(x,y,v_x,v_y) =\bar F(x,y,\xi=xv_y-yv_x)$.  Then using the method of the 
 the characteristics in
condition   $N_2F=0$,
 we can obtain the first integral:
$$
F(x,y,v_x,v_y)=\frac 12 (xv_y-yv_x)^2+\int^{x/y}\left[-\frac 1{u^3}\, f\left(\frac 1u\right)+
  u\,g\left(\frac 1u
\right)\right]\,du\,.
$$
For the Ermakov system with $f(1/u)=k$ and $g(1/u)=0$ we obtain the known Ermakov invariant
$$
F(x,y,v_x,v_y)=\frac k2\left(\frac{y}{x}\right)^2+\frac 12(xv_y-yv_x)^2
$$
We can now consider a system made up by a
  Pinney equation with two associated harmonic oscillator equations,
with associated $t$-dependent  vector field  
$$X=v_x\pd{}x+v_y\pd{}y+v_z\pd{}z+\frac{k}{y^3}
\frac{\partial}{\partial v_y}-\omega^2(t)\left(x\frac{\partial }{\partial v_x}+y\frac{\partial }{\partial v_y}+
z\pd{}{v_z}\right)
$$
which  can be expressed as $X=N_2-\omega^2(t)N_1$ where
 $N_1$ and $N_2$ are:
$$
N_1=y\frac{\partial }{\partial v_y}+x\frac{\partial }{\partial v_x}+
z\pd{}{v_z},\quad N_2=v_y\frac{\partial}{\partial y}+
\frac{1}{y^3}\frac{\partial}{\partial v_y}+v_x\frac{\partial}{\partial
  x}+v_z\frac{\partial}{\partial z},
$$ 
These vector fields generate a three-dimensional real Lie algebra
 isomorphic to   $\goth{sl}(2,\mathbb{R})$  with the
vector field $N_3$ given by 
$$
 N_3=\frac 12\left(x\frac{\partial}{\partial x}-v_x\frac{\partial}{\partial
     v_x}+y\frac{\partial}{\partial y}-v_y\frac{\partial}{\partial
     v_y}+z\frac{\partial}{\partial z}-v_z\frac{\partial}{\partial
     v_z}\right)\,.
$$

In this case there exist  three
first  integrals for  the distribution generated by these fundamental vector
fields:
 The Ermakov invariant
 $I_1$ of the subsystem involving variables $x$ and $y$, 
the Ermakov invariant $I_2$ 
of the subsystem involving variables $y$ and $z$, and finally,  
the Wronskian $W$ of the subsystem involving variables $x$ and $z$.
They are given by $W=xv_{z}-zv_{x}$,
$$I_1=\frac 12\left((yv_{x}-xv_y)^2+c\left(\frac {x}y\right)^2\right)\,,\quad
I_2=\frac 12\left((yv_{z}-zv_y)^2+c\left(\frac {z}y\right)^2\right)
\,.
$$

In terms of these three integrals we can obtain an explicit expression of $y$ in terms of $x, z$ and the integrals $I_1, I_2, W$:
$$
y=\frac {\sqrt{2}}{W}\left(I_2x^2+I_1z^2\pm\sqrt{4I_1I_2-cW^2}xz\right)^{1/2}\,.
$$

This can be interpreted as saying that there is a superposition rule allowing
us to express  the general solution of the Pinney equation in terms of two
independent solutions of the corresponding harmonic oscillator with time-dependent frequency.

\section{The reduction method and integrability criteria}

Given an equation (\ref{lieingr}) 
 on a Lie group, it may happen that the only non-vanishing coefficients
are those corresponding to a subalgebra $\goth h$ of $\goth g$ and then the equation reduces to a simpler equation on a subgroup, involving less coordinates.
An important result is that if we know a particular solution of the
 problem associated in a homogeneous space, the original solution reduces
 to one on the isotopy subgroup.

One can show that there is an action of the group 
 $\mathcal{G}$ of curves in $G$ on the set of right-invariant Lie systems in
 $G$ 
(see e.g. \cite{CRPraga} for a geometric justification), 
and we can take advantage of such
 an action for transforming a given Lie system into another simpler one. 

So, if $g(t)$ is a solution of the given Lie system and 
we choose a curve $g^\prime(t)$ in the group $G$, 
 and define 
a curve $\overline g(t)$ by $\overline g(t)=g^\prime(t)g(t)$, then 
  the new curve in 
$G$, $\overline g(t)$, determines a
 new Lie system. 
Indeed,
$$
R_{\overline g(t)^{-1}* \overline g(t)}(\dot {\overline g}(t))
=R_{g^{\prime\,-1}(t)*g^{\prime}(t)}(\dot g^\prime(t))
-\sum_{\alpha=1}^r b_\alpha(t){\rm Ad\,}(g^{\prime}(t)){\rm a}_\alpha\ , 
$$
which is similar to the original one, 
with a different right-hand side. 
Therefore, the aim is
to choose the curve $g^\prime(t)$  in such a way that 
 the new equation be simpler. 
For instance, we can choose a subgroup $H$ and 
look for a choice of $g'(t)$ such that the right hand side 
 lies in $T_e H$, and hence $\overline g(t)\in H$ for all
$t$. This can be done when we know  a 
 solution of the associated Lie system 
in $G/H$ allows us to reduce the problem to one in the subgroup $H$\cite{CarRamGra}.  

{\bf Theorem:} \ {\it Each solution of (\ref{lieingr})
on the group $G$ can be written in the 
form $g(t)=g_1(t)\,h(t)$,
where $g_1(t)$ is a curve on $G$ projecting onto a solution 
$\tilde g_1(t)$  for the left action $\lambda$ of $G$ on
the homogeneous space $G/H$
and $h(t)$ is a solution of an equation but for the
subgroup $H$,
given explicitly by}
$$
(R_{h^{-1}*h}\dot h\, )(t) 
=-{\rm Ad\,}(g_1^{-1}(t))\left(\sum_{\alpha=1}^r b_\alpha(t){\rm a}_\alpha
+(R_{g_1^{-1}*g_1}\dot g_1)(t)\right)\in {T_eH}\ .
$$

This fact is very important because one can show that
Lie systems associated 
with solvable Lie algebras are solvable by quadratures
 and therefore,
given a Lie system with an arbitrary $G$ having a solvable subgroup, we should
look for a possible transformation from the original system to one which reduces
to the subalgebra and therefore integrable by quadratures. 

By the last Theorem there always exists a curve in $G$ that transforms the initial Lie system into a new one related with solvable a Lie subgroup of $G$. Nevertheless, it can be difficult to find out a solution of the equation in $M$ that determines this transformation. Then, to be able
to obtain one is more interesting to suppose also that this transformation is a
curve in a certain subset of $G$, i.e. a one-dimensional subgrop. It would be easier
 to obtain a transformation but it may be that such a transfrormation does not
 exist. 
In summary:  {\sl 
The conditions for the existence of such a transformation of a certain form are integrability
conditions for the system}.

We could choose for showing this assertion a particular 
example: Riccati equation.
One can find in the literature a lot of integrability criteria for Riccati
equation \cite{Kamke, {Mu60}, {Stre}}, all of them particular examples of the above method \cite{CRL07d}.
We can also consider other equivalent examples as the Pinney equation, the
(generalized) Ermakov system or more relevant examples in Physics, for instance,
time-dependent harmonic oscillators.
The results obtained for one system are valid for the other; they are
essentially conditions for the equation in the group, and 
all are examples of Lie systems associated with the same Lie group:
$SL(2,\mathbb{R})$. Consider, for instance, the Riccati equation (\ref{Ricceq}).
The group $SL(2,\mathbb{R})$ contains the affine group (either the one generated by
$X_0$ and $X_1$ or the one generated by $X_1$ and $X_2$), which is SOLVABLE. Therefore, a
transformation from the given equation to one of this subgroup allows us to express the
general solution in terms of quadratures.
This happens when we know a particular solution $x_1$ of the given equation:
$x=x_1+z$, what corresponds to choose 
$$\bar g(t)=\matriz{cc}{1&-x_1\\0&1}
$$
reduces the equation  to
$ {dz}/{dt}=(2\, b_2\, x_1+b_1) z+b_2\,z^2$.
The reduction by the knowledge of two or three quadratures has also been
studied from this perspective and similarly for the Strelchenya criterion \cite{Stre}.

Each Riccati equation can be considered as a 
curve in ${\mathbb{R}}^3$ and   we can 
transform every function in $\mathbb{R}$, $x(t)$,
under an element of the group ${\GR}$
of smooth $SL(2, {\mathbb{R}})$-valued curves 
${\rm Map}({\mathbb{R}},\,SL(2,{\mathbb{R}}))$, as follows:
$$
\begin{array}{rcl}
\Theta(A,x(t))&=&{\frac{\alpha(t) x(t)+\beta(t)}{\gamma(t) x(t)+\delta(t)}}\ ,\ \ \ 
\mbox{if\ }\  x(t)\neq-{\frac{\delta(t)}{\gamma(t)}}\ ,\cr
\Theta(A,\infty)&=&{\alpha(t)}/{\gamma(t)}\ ,\ \ \ \ \Theta(A,-{\delta(t)}/{\gamma(t)})=\infty\ ,\cr &&\mbox{when}\ 
A=\matriz{cc} {{\alpha(t)}&{\beta(t)}\\{\gamma(t)}&{\delta(t)}}\,\in{\cal G}\
.\end{array}
$$
The image $x'(t)=\Theta({\bar A}(t),x(t))$ of 
 a curve $x(t)$ solution of the given Riccati equation
satisfies a new Riccati equation with the coefficients $b'_2,b'_1, b'_0$: 
$$
\begin{array}{rcl}
b'_2&=&{\delta}^2\,b_2-\delta\gamma\,b_1+{\gamma}^2\,b_0+\gamma{\dot{\delta}}-\delta \dot{\gamma}\ ,
\nonumber \cr
b'_1&=&-2\,\beta\delta\,b_2+(\alpha\delta+\beta\gamma)\,b_1-2\,\alpha\gamma\,b_0   
       +\delta\dot{\alpha}-\alpha \dot{\delta}+\beta \dot{\gamma}-\gamma \dot{\beta}\ ,   \cr
b'_0&=&{\beta}^2\,b_2-\alpha\beta\,b_1+{\alpha}^2\,b_0+\alpha\dot{\beta}-\beta\dot{\alpha} \ .
 \nonumber 
\end{array}
$$
This expression defines an affine action of the group  ${\GR}$ on the set of 
Riccati equations or analogous Lie systems.

 Lie systems in $SL(2,\mathbb{R}) $ defined by a constant curve,
${\rm a}(t)=  \sum_{\alpha=0}^2 c_\alpha {\rm a}_\alpha$,
are integrable and  the same happens for 
curves of the form ${\rm a}(t)= D(t)\left( \sum_{\alpha=0}^2 c_\alpha {\rm
    a}_\alpha\right)$, where $D$ is an arbitrary function, because a time reparametrisation reduces the problem to
the previous one, i.e. the system is essentially a Lie system on a one-dimensional Lie group.

We can prove the following theorem which is valid  for both
Riccati equation and  any other Lie system with Lie group $SL(2,\mathbb{R})$: 

{\bf Theorem:} \ {\it The necessary and sufficient conditions
for the existence of a  transformation:
$y'=G(t)y$, i.e. $ \bar A(t)=\matriz{cc}{\alpha(t)&0\\0&\alpha^{-1}(t)}$,
 relating the Riccati equation (\ref{Ricceq}) with 
(for $b_0b_2\ne 0$,
with an integrable one given by
\begin{equation}
\frac{dy'}{dt}=D(t)(c_0+c_1y'+c_2y'^2)\,,\label{solvaRicc}
\end{equation}
where $c_i$ are real numbers,  $c_i\in \mathbb{R}$, is that  $c_0c_2\ne 0$,  
and:
$$
D^2(t)c_0c_2=b_0(t)b_2(t),\qquad \sqrt{\frac{c_0c_2}{b_0(t)b_2(t)}}\left(b_1(t)+\frac{1}{2}\left(\frac{\dot b_2(t)}{b_2(t)}-\frac{\dot b_0(t)}{b_0(t)}\right)\right)=c_1.
$$
The unique transformation is then $y'=(b_2(t)c_0)^{1/2}(b_0(t)c_2)^{-1/2}\,y$.}

As a consequence, given (\ref{solvaRicc})
if there are constants $K,L$ such that 
$$
 \sqrt{\frac{L}{b_0(t)b_2(t)}}\left({b_1(t)+\frac{1}{2}\left(\frac{\dot b_2(t)}{b_2(t)}-\frac{\dot b_0(t)}{b_0(t)}\right)}\right)=K\,
$$
then there exists a 
 time-dependent linear change of variables  transforming the given equation into
 the solvable  Riccati equation (\ref{solvaRicc}) with 
${c_1}=K, c_0\,c_2=L$
and $D(t)$ is given as above.

The existence of such constant $K$ can be considered a sufficient condition for
integrability of the given Riccati equation or the corresponding Milne--Pinney
equation.



\begin{thebibliography}{50}

\bibitem{CarRam}
J.~F. Cari\~nena and  A. Ramos,
{\em Int. J. Mod. Phys.}  {\bf A 14}, 1935 (1999).

\bibitem{Pi50}{E. Pinney},
{\em   Proc. Am. Math. Soc.} {\bf 1}, 681 (1950).

\bibitem
{LS}   S. Lie and G. Scheffers,
{\it Vorlesungen \"uber continuierliche Gruppen mit geometrischen
 und anderen Anwendungen}, Edited and revised by G. Scheffers,
 (Teubner, Leipzig, 1893).

\bibitem{CGM00}
J.~F. Cari\~nena,  J. Grabowski  and G. Marmo,
{\it Lie--Scheffers systems: a geometric approach},
(Bibliopolis, Napoli, 2000).

\bibitem{CGM07}
J.~F. Cari\~nena,J. Grabowski and  G. Marmo, 
Rep. Math. Phys. {\bf 60}, 237--58 (2007)

\bibitem{CRPraga}
J.~F. Cari\~nena  and  A. Ramos, 
 in {\it 9$^{\text
th}$ Int. Conf. Diff. Geom and Appl.}, p. 437--452    (2004), J.
Bures {\sl et al.}  Eds.,   (Matfyzpress, Praga,  2005).

\bibitem{CarRamGra}
J.~F. Cari\~nena, J. Grabowski and  A. Ramos,
 {\em  Acta Appl.  Math.} {\bf 66}, 67 (2001).

\bibitem{Kamke}
E. Kamke,
{\it Differentialgleichungen: L\"osungsmethoden und L\"osungen},
(Akademische Verlagsgeselischaft, Leipzig, 1959).

\bibitem{Mu60}  G.~M. Murphy,
{\it Ordinary differential equations and their solutions},
(Van Nostrand, New York, 1960).

\bibitem{Stre} V.~M. Strelchenya,
{\em J. Phys. A: Math. Gen.} {\bf 24}, 4965 (1991).

\bibitem{CRL07d}
J.~F. Cari\~nena, A. Ramos and J. de Lucas,
{\em Electron. J. Diff. Eqns.} {\bf  122},  1 (2007).





\end{thebibliography}
\end{document}